# Pump-probe nuclear spin relaxation study of the quantum Hall ferromagnet at filling factor $\nu = 2$

K. F. Yang,[1,7] M. M. Uddin,[2,3,7] K. Nagase,[2] T. D. Mishima,[4] M. B. Santos,[4] Y. Hirayama,[2,5] Z. N. Yang,[6] and H. W. Liu[1,8]

[1]State Key Lab of Superhard Materials, Jilin University, Changchun 130012, P. R. China

[2]Department of Physics, Tohoku University, Sendai, Miyagi 980-8578, Japan

[3]Department of Physics, Chittagong University of Engineering & Technology, Chattogram-4349, Bangladesh

[4]Homer L. Dodge Department of Physics and Astronomy, University of Oklahoma, 440 West Brooks, Norman, OK 73019-2061, USA

[5]Center for Science and Innovation in Spintronics (Core Research Cluster), Tohoku University, Sendai 980-8577, Japan

[6]College of Physics, Jilin University, Changchun 130012, P. R. China

[7]These authors contributed equally to this work.

[8]Author to whom any correspondence should be addressed.

E-mail: hwliu@jlu.edu.cn

**Abstract.** The nuclear spin-lattice relaxation time $T_1$ of the $\nu = 2$ quantum Hall ferromagnet (QHF) formed in a gate-controlled InSb two-dimensional electron gas has been characterized using a pump-probe technique. In contrast to a long $T_1$ of quantum Hall states around $\nu = 1$ that possesses a Korringa-type temperature dependence, the temperature-independent short $T_1$ of the $\nu = 2$ QHF suggests the presence of low energy collective spin excitations in a domain wall. Furthermore, $T_1$ of this ferromagnetic state is also found to be filling- and current-independent. The interpretation of these results as compared to the $T_1$ properties of other QHFs is discussed in terms of the domain wall skyrmion, which will lead to a better understanding of the QHF.

## 1. Introduction

The quantum Hall ferromagnet (QHF) formed at two energetically-degenerate spin-resolved Landau levels (LLs) in a two-dimensional electron gas (2DEG) has provided an ideal system for understanding itinerant electron ferromagnetism, spin interactions, and domain dynamics [1-13]. In particular, a resistively detected nuclear magnetic resonance (RDNMR) technique developed in the QHF of GaAs 2DEGs at filling factor $\nu = 2/3$ (corresponding to a composite-fermion (CF) filling factor $\nu_{CF} = 2$) has been widely used to investigate the dynamic nuclear polarization (DNP) in semiconductors [14, 15], to coherently control the nuclear spins in the 2DEG [16], and to discover exotic electron phases in quantum Hall systems [17-19]. Furthermore, this highly-sensitive technique combined with the nuclear spin-lattice relaxation time $T_1$ measurement [7, 8] as a unique probe of low-frequency spin fluctuations can be applied to investigate the domain-wall (DW) structures of the QHF that are still poorly characterized, which will lead to a more comprehensive understanding of the QHF. Up to now, the $T_1$ results of the QHF formed at $\nu = 2/3$ [20] and at integers $\nu$ in a two-subband 2DEG (hereafter called two-subband QHF) [21, 22] are suggestive of low energy DW excitations but exhibit different properties. Note that intricate CF-CF interactions at fractional $\nu$ (different from electron-electron interactions at integer $\nu$) [23] and an additional degree of freedom associated with the subband index [24] may complicate the interpretation of these differences.

More recently, we have performed the RDNMR measurement of the QHF formed at $\nu = 2$ of the InSb 2DEG [25, 26]. Because the $\nu = 2$ QHF including only the two lowest LLs can be described by a single-particle picture, it is more appropriate for investigating the DW dynamics of the QHF. Although our preliminary results show that $T_1$ in this QHF is independent of temperature ($T$), it is unclear whether such a $T_1$ property is unique or not because the $T_1$ measurement of the non-QHF states by means of a pump-probe technique is not available due to the difficulty in fabricating the InSb 2DEG with gate control. With our recent success in the fabrication of the gate-controlled InSb 2DEG [27, 28] it is of immense interest for us to systematically investigate the $v = 2$ QHF using the pump-probe $T_1$ measurement. Here we report on the $T_1$ characterization of the $v = 2$ QHF and make a comparison with that of other QHFs.

## 2. Samples and methods

The pump-probe $T_1$ measurement in this study was carried out on two gate-controlled InSb 2DEGs; one with an $Al_{0.1}In_{0.9}Sb$ surface layer (Sample 1) [28] and the other with an InSb surface layer (Sample 2) [27]. Both samples were patterned into a Hall bar with a length of $L = 100$ μm and a width of $W = 30$ μm. A low noise preamplifier (Stanford Research Systems, Model SR560) was used for DC magnetotransport and RDNMR measurements. A small RF field ~ μT (continuous wave mode) matching the NMR resonance frequency of individual nuclei was generated by a single turn coil surrounding the sample (right panel, figure 1(a)). The RDNMR signal was defined by the change in longitudinal resistance $R_{xx}$ of the Hall bar at the resonance condition [29]. All the measurements were performed using a dilution refrigerator with an *in situ* rotator at a base temperature of $T = 30$ mK (unless otherwise noted). The tilt angle $\theta$ between the sample normal and the direction of the total magnetic field ($B$) was determined by the Hall measurement at low fields.

The ratio of Zeeman ($E_z \propto B$) to cyclotron ($E_c \propto B_{perp}$, the perpendicular component of $B$) splittings can be tuned by $\theta$ to bring the LLs with opposite spins into degeneracy (right panel, figure 1(a)). In a single-particle picture, a first-order spin phase transition from an unpolarized state ($P = 0$) to a fully polarized state ($P = 1$) occurs when the spin-down (↓) LL with orbital number $n = 0$ and the spin-up (↑) LL with $n = 1$ intersect, resulting in the formation of the $v = 2$ QHF [4]. However, strong electron-electron interactions at the LL intersection will cause this ferromagnetic state to occur *before* the single-particle levels cross, where the energy difference ($E_c - E_z$) is compensated by the exchange energy $E_{ex}$ [3]. The evolution of these two LLs is mapped by the position of magnetoresistance peaks on both sides of $v = 2$ as a function of $\theta$, and in between an emerging peak characterizing the $v = 2$ QHF is found to move in a certain range of $\theta$ due to the $B_{\mathrm{perp}}$-dependent $E_{\mathrm{ex}}$ (see Supplementary Material). Charge transport through the DW is expected to be responsible for this peak [29], around which the RDNMR signal is detected by aid of the current-induced DNP due to the degeneracy of DW states [30]. Because gate control of the electron density $n_{\mathrm{s}}$ (or $v$) is required for the pump-probe measurement, the dependence of magnetoresistance on gate voltage $V_{\mathrm{g}}$ at a given $\theta$ with strong RDNMR signals has to be examined (Supplementary Fig. S2), from which the $V_{\mathrm{g}}$ dependence of $R_{\mathrm{xx}}$ used for the pump-probe measurement is obtained. Figure 1(a) shows the dependence of $R_{\mathrm{xx}}$ and transverse (Hall) resistance $R_{\mathrm{xy}}$ on $V_{\mathrm{g}}$ (or $v$) of Sample 1. As we know, the QH effect is characterized by a plateau in $R_{\mathrm{xy}}$ with zero $R_{\mathrm{xx}}$ and by a plateau-to-plateau transition with a concurrently developed $R_{\mathrm{xx}}$ peak. However, the $R_{\mathrm{xx}}$ peak at $V_{\mathrm{g}} \sim -0.45$ V in Fig. 1(a) is found to reach the deep side of the $v = 2$ plateau with nonzero $R_{\mathrm{xx}}$ minimum, which is attributed to the formation of the $v = 2$ QHF (see Supplementary Material for a detailed description). This conclusion is further supported by the fact that the RDNMR signal is observed in the region between $V_{\mathrm{g}} = -0.3$ V and $V_{\mathrm{g}} = -0.5$ V (hereafter called QHF region)(data not shown). The pump-probe $T_1$ measurement was performed at $V_{\mathrm{g}} = -0.4$ V (corresponding to $v \sim 1.73$ as defined by $v_{\mathrm{p}}$) as shown in figure 1(b). A pump current of 1 µA was applied to polarize the nuclei as indicated by an exponential increase in $R_{\mathrm{xx}}$. $R_{\mathrm{xx}}$ became saturated ($R_{\mathrm{xx}}^{\mathrm{sat}}$) after a time period of $\tau_{\mathrm{p}}$ (Step I). The current was then switched off and $v_{\mathrm{p}}$ was immediately tuned to the probe filling factor $v_{\mathrm{d}}$ by $V_{\mathrm{g}}$, where the polarized nuclei were expected to depolarize due to the electron-nuclear hyperfine interaction (Step II). After a time period of $\tau_{\mathrm{d}}$, the sample was restored to the original pump condition and the nuclei were polarized again as indicated by the change in $R_{\mathrm{xx}}$ with respect to $R_{\mathrm{xx}}^{\mathrm{sat}}$ ($\Delta R_{\mathrm{xx}}$, Step III). The plot of $\Delta R_{\mathrm{xx}}$ vs. $\tau_{\mathrm{d}}$ (figure 1(c)) was obtained by repeating this pump-probe-pump sequence at a certain $v_{\mathrm{d}}$, and a fit to the data gave $T_1$.

## 3. Experimental results and discussion

Figure 2(a) shows that $T_1$ in the QHF region is constant (~ 45 s) and is at least one order of magnitude smaller than that at $v_d = 1$ (~ 540 s). Furthermore, $T_1$ near $v_d = 1$ is found to be filling dependent, resembling the results of the GaAs 2DEG where a single-particle model taking account of the disorder-induced LL broadening and the exchange-enhanced activation energy gap accounts for the nuclear spin relaxation [31, 32]. Note that the skyrmion is not expected to contribute to $T_1$ near $v_d = 1$ because it is not available due to large Zeeman splitting of the InSb 2DEG [33, 34]. It should be pointed out that $T_1$ at $V_g = -0.4$ V ($v_d \sim 1.73$) is the same as that obtained from the time dependence of $R_{xx}$ in Step I of figure 1(b), suggesting a current-independent $T_1$. Similar results are also obtained in Sample 2 (figure 2(b)). Although layer structures and transport parameters of the two samples are different, the results of $T_1$ in the QHF region are essentially the same. These findings cannot be explained by the single-particle model.

A further study was carried out on the temperature dependence of the nuclear spin-lattice relaxation rate $T_1^{-1}$ of Sample 1 as shown in figure 3. In striking contrast to a linearly $T$-dependent $T_1^{-1}$ around $v_d = 1$ that suggests a Korringa-type relaxation process, $T_1^{-1}$ in the QHF region is found to be independent of $T$. The temperature-independent $T_1^{-1}$ is also observed in the two-subband QHF [21] and implicitly present in the $v = 2/3$ QHF [35]. Therefore, it is concluded that the $T$-independent $T_1^{-1}$ is a common feature of the QHF characterized by the RDNMR measurement. Besides this, the current study also demonstrates that there are still distinct differences in the $T_1$ properties of various QHFs. The $T_1$ measurement of the $v = 2/3$ QHF gives a current-dependent relaxation time constant between 180 s and 1350 s [20], while that of the $v = 2$ QHF is relatively short (~ 45 s) and current-independent. Furthermore, $T_1$ of the two-subband QHF is filling dependent [22] as contrasted with the $v$–independent $T_1$ of the $v = 2$ QHF. Note here that the magnetotransport data are indicative of thermally activated transport in both the $v = 2/3$ and two-subband QHFs of the GaAs 2DEG [20, 22]. A sharp reduction in the activation energy gap at the phase transition suggests the presence of skyrmion excitations in the DW that dominate the nuclear spin relaxation and thus $T_1$ [20]. Because the precise spin textures depend on the strength of the Zeeman coupling ($\propto B$) [34], their size and energy are expected to vary with $\nu$ ($\propto 1/B$) that accounts for the filling-dependent $T_1$. Furthermore, a large current will increase the number of these excitations as suggested by the optically detected magnetic resonance imaging (ODMRI) of the $\nu =$ 2/3 QHF [13] and thus enhance the relaxation efficiency, resulting in a short $T_1$. In fact, the skyrmion trapped in the DW (called DW skyrmion [36-38]) have been theoretically predicted to exist in the $\nu$ = 1 easy-axis QHF with domain structures [39] and the $\nu = 2$ QHF [6]. In contrast to the extensively studied quantum Hall skyrmion near the $\nu = 1$ isotropic QHF without domain structures that has a spherically symmetric (non-chiral) hedgehog configuration in order-parameter space, the DW skyrmion has a combed (chiral) hedgehog configuration by performing a $\pi/2$ rotation about the z-axis towards the north pole [40]. The in-plane (xy) component of a unit vector of spins gains a $2\pi$ phase within the size length $\xi$ of this excitation along the DW [6,37]. The low-frequency fluctuations in the xy spin component of the DW skyrmion (probably in a liquid state [41]) will accelerate the relaxation and thus result in a short $T_1$ [42]. Following the above discussion, the results of this study have been interpreted in terms of the DW skyrmion. We note that the variable-range hopping (VRH) rather than the thermally activated transport is dominant in the $\nu = 2$ QHF of the InSb 2DEG [26], suggesting that the activation energy gap is much larger than the thermal energy (see Supplementary Material). This ensures that the low-energy DW skyrmion dominates the nuclear spin relaxation over the range of temperatures investigated [43], showing the $T$ independence of $T_1$. Although the VRH transport does not provide additional information of the DW skyrmion, the $T_1$ results suggest the size and number of such an excitation to be independent of both $\nu$ and current based on the above discussion. In fact, the current density typically used in the RDNMR measurement of the InSb 2DEG is at least ten times that of the GaAs 2DEG. It is shown in the ODMRI of the $\nu = 2/3$ QHF that striped DW structures occur at large current [13]. In this case, the density and size of the DW skyrmion is probably determined by the distribution and morphology of the DW that is responsible for the current- and $\nu$–independent $T_1$.

Finally, we discuss the effect of spin-orbit coupling (SOC) on $T_1$. The electron transport between domains with opposite spin polarization is known to be accompanied by the spin-flip process that is usually mediated by electron-nuclear hyperfine interaction or SOC in order to conserve the angular momentum. Although it is evident from the above results and discussions that the hyperfine interaction dominates the spin flip, the effect of SOC cannot be completely excluded. In fact, the possible role of SOC in the nuclear spin relaxation has been investigated using the $\nu = 2/3$ QHF [9]; the structure inversion asymmetry-induced Rashba SOC is believed to facilitate the spin flip and thus to enhance the nuclear spin relaxation rate by fluctuating the hyperfine field. It is worth noting that theoretical analyses of the coupling of the LLs due to the Rashba or Dresselhaus SOC are restricted to the electron rather than the CF system [44,45]. Therefore, the $\nu = 2$ QHF is more appropriate for investigating the interplay between the hyperfine interaction and SOC in $T_1$. A strong SOC in InSb [46] is helpful to examine the effect of different types of SOC on $T_1$ by virtue of both in-situ high pressure and gate control approaches, and such a study is currently underway in our laboratory.

## 4. Conclusions

The $T_1$ properties of the $\nu = 2$ QHF in the InSb 2DEG that have been characterized in this study，combined with those of the $\nu = 2/3$ and two-subband QHFs in the GaAs 2DEG, demonstrate that the $T$-independent $T_1$ is a common feature of the QHF. We discuss the filling, current, and temperature dependence of $T_1$ of these QHFs in connection with the theoretically predicted DW skyrmion. Local probe of DW electronic structures of the $\nu = 2$ QHF is a major step towards the experimental discovery of the DW skyrmion that is distinctly different [37] from the previously known quantum

Hall skyrmion and magnetic skyrmion [47].

**Acknowledgments**

This work was supported in part by the grants Program for New Century Excellent Talents of University in China (to H. W. L.), National Natural Science Foundation of China (No. 11704144, to K. F. Y.), Jilin Natural Science Foundation (No. 20180101286JC, to K. F. Y.), Fundamental Research Funds for the Central Universities (to K. F. Y.), S-SDC fellowship in Tohoku University (to M. M. U), JST-ERATO, KAKENHI (Nos. 15H05867 and 18H01811, to Y.H.), CSRN in Tohoku University (to Y. H.), and GP-Spin in Tohoku University (to K. N. and Y. H.).

**Figure 1.** (a) Longitudinal resistance $R_{xx}$ and Hall resistance $R_{xy}$ as a function of gate voltage $V_g$ (or filling factor $\nu$) of Sample 1 at $T = 30$ mK, $\theta = 65°$, $B = 15$ T, and a DC current of $I = 35$ nA. The right panel schematically illustrates the measurement setup and the evolution of the LLs indexed by n with increasing $\theta$. The Zeeman and cyclotron energy gaps are denoted by $E_z$ and $E_c$, respectively, and a downward (upward) arrow is for spin-down (spin-up). (b) Plot of $R_{xx}$ vs. time for the time sequence of the pump-probe nuclear spin-lattice relaxation $T_1$ measurement. Step I: polarization of nuclei at $\nu_p \sim 1.73$ ($V_g = -0.4$ V) with a pump current of $I = 1$ µA until $R_{xx}$ reaches a saturated value $R_{xx}^{sat}$ after a duration time $\tau_p$; Step II: depolarization of nuclei at the probe filling factor $\nu_d$ in the absence of current for a certain time $\tau_d$; Step III: repolarization of nuclei under the condition of Step I. The change in $R_{xx}$ with respect to $R_{xx}^{sat}$ gives $\Delta R_{xx}$. (c) $\Delta R_{xx}$ vs. $\tau_d$ at $\nu_d = 1$ ($V_g = -0.63$ V). The solid line is an exponential fit to the data.

**Figure 2.** $T_1$ (circles with a thin line as a guide to the eye) and $R_{xx}$ (thick line) of Sample 1 (a) and Sample 2 (b) as a function of $V_g$ (or $\nu_d$). Note that the range of gate-controlled electron density $n_s$ for the two samples is different; that for Sample 1 (Sample 2) varies from 4.05 (4.2) × 10$^{15}$ m$^{-2}$ to 1.31 (2.5) × 10$^{15}$ m$^{-2}$ as $V_g$ is tuned from 0 V to -0.7 (-4) V. In addition, the low-temperature electron mobility $\mu$ of Sample 1(Sample 2) is about 4.5 (16) m$^2$/Vs measured at $n_s \sim 2.5 \times 10^{15}$ m$^{-2}$. These results lead to the difference in $\nu_p$ and $\nu_d$ for the pump- probe measurement of the two samples; $\nu_p \sim$ 1.73 ($V_g = -0.4$ V) for Sample 1 and $\nu_p \sim 1.86$ ($V_g = -4$ V) for Sample 2. The current used for the $T_1$ and $R_{xx}$ measurements of Sample 1 (Sample 2) are 1 µA (760 nA) and 35 nA, respectively.

**Figure 3.** Temperature ($T$) dependence of the nuclear spin-lattice relaxation rate $T_1^{-1}$ for Sample 1 at different $V_g$ (or $\nu_d$). The solid line is a fit to the data using the Korringa law: $T_1 T = $ constant. The data are deviated from the fit at $T = 30$ mK where the nuclear spin diffusion is expected to make an additional contribution to $T_1^{-1}$ [31]. The dashed line is a guide to the eye.

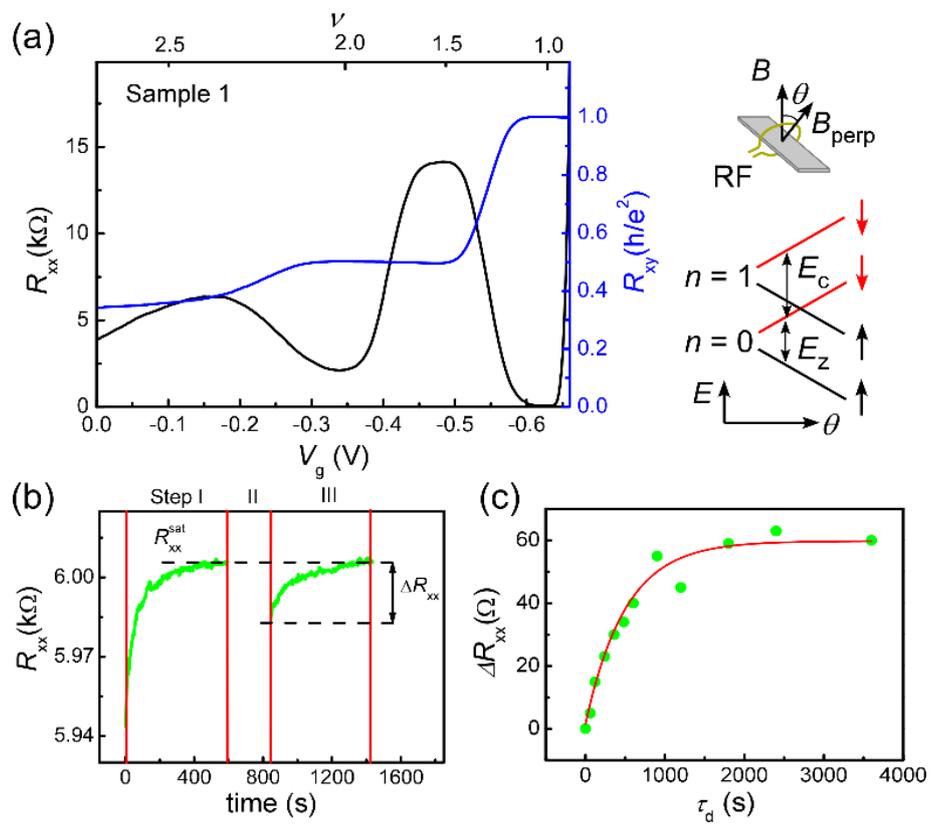

Figure 1

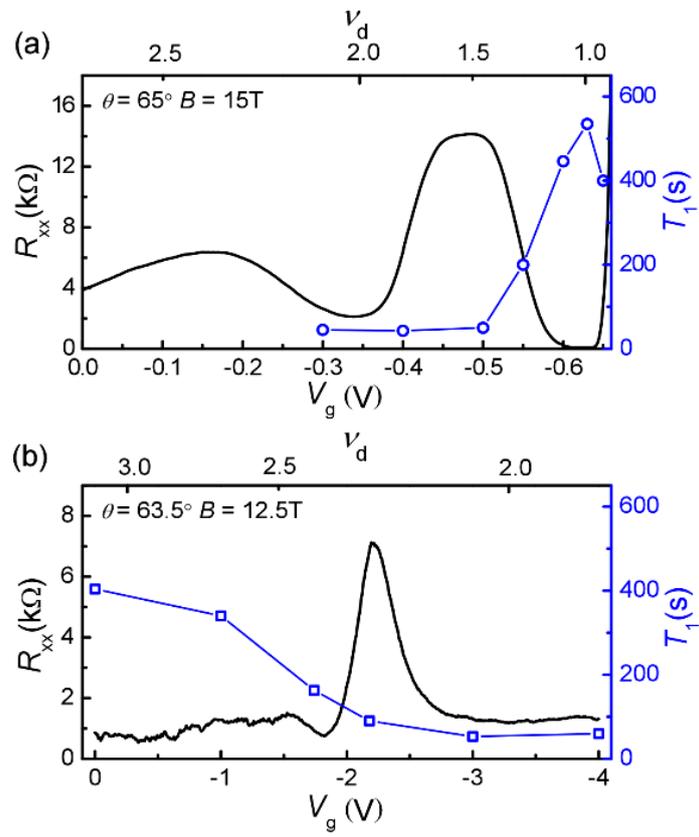

Figure 2

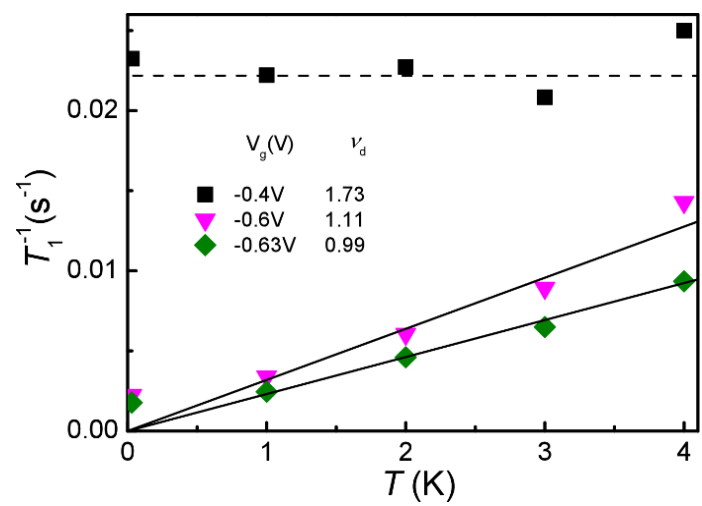

Figure 3

Supplementary Materials

This supplementary material has been prepared to describe the experimental procedure used to collect the magnetoresistance data in Fig. 1(a) and Fig. 2(b). As one example, results of Sample 2 are reported here. The electron density $n_s$ of Sample 2 reaches a saturated value of ~2.5 × $10^{15}$ m$^{-2}$ at gate voltage $V_g$ = -4 V [S1] that is low enough to allow the $\nu = 2$ quantum Hall ferromagnet (QHF) to occur below the maximum magnetic field of 15 T used in our cryostat. It is shown in Fig. S1 that the angle dependence of two magnetoresistance peaks on both sides of $\nu = 2$ describing the evolution of Landau levels (LLs) exhibits the anticrossing behavior. The two branches of this anticrossing structure correspond to the spin-down (↓) LL with orbital number $n = 0$ and the spin-up (↑) LL with $n = 1$ (right panel of Fig. 1(a), main text) rather than to their coherent superposition states [S2], and in between a peak characterizing the QHF (hereafter called QHF peak) occurs. This emerging peak moves in a narrow range of angles between $\theta = 63°$ and $\theta = 65°$ (dashed oval) due to a linear dependence of the exchange energy $E_{ex}$ on the perpendicular component of $B$ ($B_{perp}$) [S3]. Note that the QHF peak is not clearly identified in Fig. S1 due to its broadening and overlap with neighboring peaks, but distinct in a sample without gate [S4]. As a reference, we estimate the values of $E_{ex}$ between 8.6 meV and 13 meV that are linear in $B_{perp}$, and the activation energy gap that is much larger than 4 meV as suggested by the variable range hopping mechanism of the QHF peak in the sample of [S4]. The gate effect on the broadening of the QHF peak requires further study.

Furthermore, this QHF region is also confirmed by the highly-sensitive RDNMR measurement as done in [S4] (data not shown).

Figure S2 shows the $V_g$-dependent magnetoresistance at $= 63.5\,°$ where a strong RDNMR signal of the QHF is observed. As $|V_g|$ decreases, the QHF peak is found to gradually separate from the merged LL peak as indicated by a dashed line. A measure taken along the solid line at $B = 12.5$ T gives the $V_g$-$R_{xx}$ plot of Fig. 2(b) in the main text. Note that strong RDNMR signals are obtained at the tail of rather than on top of the QHF peak, where a shift of the peak after the DNP is relatively large that leads to a large change in $R_{xx}$. It is seen from Fig. S2 that data between $V_g = -3$V and $V_g = -4$V in Fig. 2(b) do not correspond to the $R_{xx}$ minimum of the quantum Hall state but to $R_{xx}$ related to the QHF. The $V_g$ dependence of the concurrently measured $R_{xx}$ and $R_{xy}$ of Sample 1 in Fig. 1(a) is also obtained following the above procedure, where a shift of the $R_{xx}$ peak to the deep side of the $v = 2$ Hall resistance plateau with nonzero $R_{xx}$ minimum is caused by the formation of the QHF.

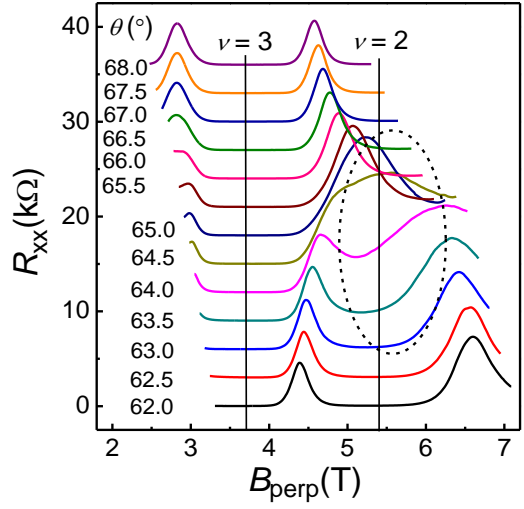

**Figure S1.** Tilt angle ($\theta$) dependence of longitudinal resistance $R_{xx}$ versus the perpendicular component of $B$ ($B_{perp} = B\cos\theta$) of Sample 2 at a DC current of $I = 35$ nA, $V_g = -4$ V, and $T = 30$ mK. A solid line indicates the integer filling factor $v = \dfrac{n_s h}{eB_{perp}}$ (where $h$ is Planck's constant and $e$ the electron charge). The QHF region is highlighted by a dashed oval. Note that the $R_{xx}$ peak at $v < 2$ is missed at large $\theta$ because the field strength required therein is greater than 15 T.

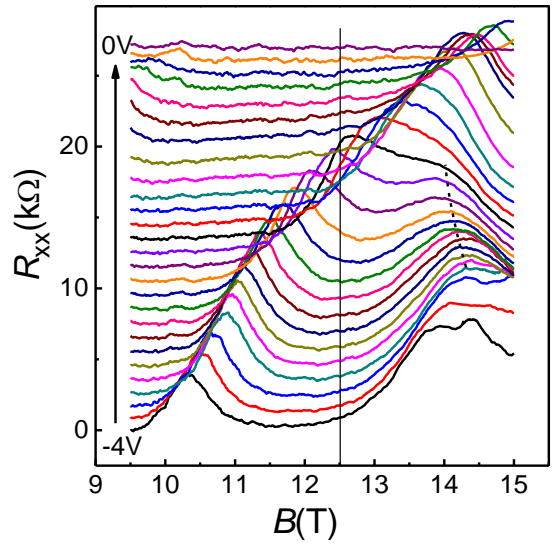

**Figure S2.** $R_{xx}$ versus $B$ for different $V_g$ at $\theta = 63.5°$, $I = 35$ nA, and $T = 30$ mK. The dashed line guides the QHF peak, and data taken along the solid line give the plot of the $V_g$ dependence of $R_{xx}$ used for the pump-probe measurement.